 \documentclass[final,5p,times,twocolumn]{elsarticle}

\usepackage{amssymb}
\usepackage{lipsum}
\usepackage{graphicx}%
\usepackage{multirow}%
\usepackage{amsmath,amssymb,amsfonts}%
\usepackage{amsthm}%
\usepackage{mathrsfs}%
\usepackage[title]{appendix}%
\usepackage{xcolor}%
\usepackage{textcomp}%
\usepackage{manyfoot}%
\usepackage{booktabs}%
\usepackage{algorithm}%
\usepackage{algorithmicx}%
\usepackage{algpseudocode}%
\usepackage{listings}%
\usepackage{orcidlink}%
\usepackage{dcolumn}%

\journal{arXiv}

\begin{document}

\begin{frontmatter}

\title{Non-equilibrium Noise in V-Shaped Linear Well Profiles}

\author[first]{Noah M. MacKay\,\orcidlink{0000-0001-6625-2321}}
\affiliation[first]{organization={Universität Potsdam, Institute of Physics and Astronomy},
            addressline={Karl-Liebknecht-Straße 24/25}, 
           postcode={14476},  
           city={Potsdam},
            country={Germany}}

\begin{abstract}
Non-equilibrium noise is characterized as noise realizations where external agitations disrupt the harmonic equilibrium of Brownian motion. Excitations in a particle's random walk into a so-called L\'evy flight changes the distribution of the noise from Gaussian to the fat-tailed L\'evy distribution. Generalization between Gaussian and L\'evy distributions is the $\alpha$-stability distribution, where $1<\alpha\leq2$. In this study, the $\alpha$-stability distributed noise is subjugated into the Langevin and fractional Fokker--Planck equations that correspond to a V-shaped linear potential well $V(x)=F|x|$. From these equations, an Euler scheme for computational simulation via iterations is presented, and a probability density function that is normalizable under any $\alpha\in(1,2]$ is obtained. This study is focused more on the theoretical framework of non-equilibrium noise in V-shaped linear well profiles, which is intended to be applied to systems known to exhibit self-organized criticality.
\end{abstract}


\end{frontmatter}




\section{Introduction}

Self-organized criticality (SOC) is a process in which a system that is progressively becoming unstable over time experiences a sudden, chaotic moment of self-regularity. Bak, Tang and Wiesenfeld \cite{Bak1987}, and their sandpile model \cite{Held1990}, considered these sudden, chaotic moments to be attributable to avalanches. These avalanches in various self-organized critical states resemble $1/f$ noise: non-equilibrium noise in which low frequencies are displayed in a power law behavior $f^{-\alpha}$. Here, $\alpha$ is the noise stability factor that has a range of $1<\alpha\leq2$: $\alpha\simeq1$ resembles L\'evy-Cauchy (in this report shortened as ``L\'evy'') noise, and $\alpha=2$ resembles Gaussian noise.

Noise analysis in physics is straightforwardly the analysis of energetic particles within a potential well drawn by $V(x)$. In both quantum-mechanical gyrations and thermally induced excitations, energetic particles induce Brownian noise via collisions with their nearest neighbors. Since the directions are random, the distances covered have a Gaussian distribution: $\rho(x)\propto\exp[-x^2/\sigma^2]$. For small $x$, $\rho(x)\rightarrow1$; for very large $x$, $\rho(x)\rightarrow0$. Based on the random walk of the particle, outliers due to large steps are rare for the Gaussian case. However, external agitations acting on the randomly walking particle ``kick'' it into a large step, causing a so-called L\'evy flight. These kicks induce non-equilibrium L\'evy noise in the Brownian bath, which introduces outliers to the Gaussian distribution and changes it into the $\alpha$-stable distribution: $\rho(x)\sim|x|^{-\alpha}$. This $\alpha$-stable distribution is analogous to the $1/f$ noise power law. Therefore, SOC systems have an inherent $\alpha$-stable distribution, treating the noise as a L\'evy bath with random flights in the Brownian system with neglected dampening.

Analytically, a particle's random walk is expressed by the Langevin equation \cite{lang}, which is a classically framed equation of motion in which the forces are calculated from the potential gradient and from an $\alpha$-dependent noise generator. In SOC systems, where equilibrium is chaotic and sudden, our choice of potential well is a sharp-cornered V-shape well drawn by $V(x)\propto|x|$. To extract a distribution function of the noise signal, and with it a probability density function (pdf) describing the likelihood of achieving equilibrium, under a certain value of $\alpha$, one can define a corresponding (fractional) Fokker--Planck equation \cite{fp, Jespersen1999}. In this short study, I aim to define an $\alpha$-stability distribution for a V-shaped linear well and normalize it under any value of $\alpha\in(1,2]$. In addition, I will describe how one can computationally simulate noise signals under any $\alpha$ stability, down-slide force and noise amplitude using the \textit{Mathematica} computation software.

SOC has many applications outside of physics, such as economic fluctuations \cite{Scheinkman1994}, neurobiology \cite{LinkenkaerHausen2001, Beggs2003, Chialvo2004, Dmitriev2021} and the stability of democracies \cite{Wiesner2019}. SOC in physics is seldom in comparison, however applications include geophysics \cite{Smalley1985, Hatamian1996, Smyth2019} and the observation of x-ray bursts from black holes \cite{bak}. Linear potential well profiles in physics include the classical gravitational potential inside a solid sphere, $V(x)=\kappa |x|$ ($\kappa$ is the surface gravity), and part of the QCD potential for quark confinement, $V_{\mathrm{QCD}}(x)=-\phi|x|^{-1}+F_0|x|$, where $\phi=4\hbar c\alpha_s/3$ is the leading term coefficient and $F_0$ is a parameter fitted from the corresponding data \cite{griffiths}. Further applications of SOC in physics are open to exploration.

\section{Methods} \label{methods}

\begin{figure} [h!]
\centering
\includegraphics[width=87mm]{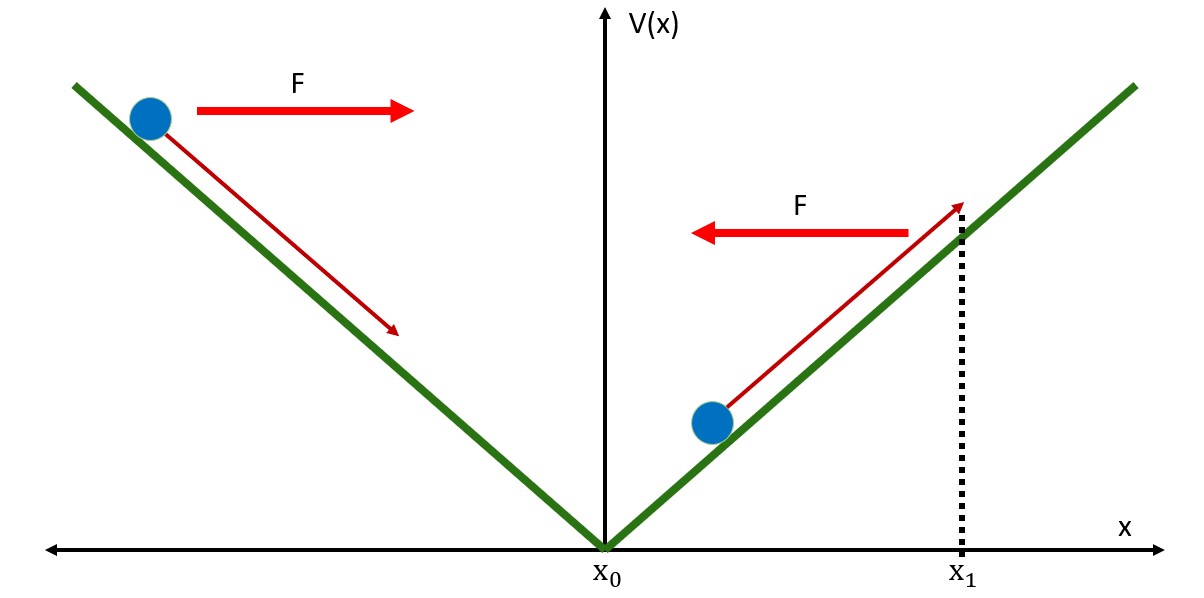}
\caption{\label{fig:well} A particle (blue marble) is randomly kicked up the V-shaped linear well. The L\'evy flight along the slope projects a horizontal ``kick length" $\lambda=x_1-x_0$. The particle then slides back down with a constant force $F$.}
\end{figure}

If one were to track just one particle in the L\'evy bath, it would be tracked as though the particle is kicked up either side of the potential well $V(x)=F|x|$ and driven back down to the equilibrium point, i.e., the node of the well. Changes in this potential over the distance covered resembles the gradient of the potential; its negative-value counterpart defines a force driving a kicked particle back to equilibirum. For a V-shaped well, this down-slide force is constant: $F(x)\equiv-dV(x)/dx=F$, as shown in Figure \ref{fig:well}.

\subsection{The Langevin Equation}

The associated Langevin equation for a particle in the linear potential well, i.e., $V(x)=F|x|$ under non-equilibrium noise, i.e., $1<\alpha\leq2$ and neglected dampening is 
\begin{equation} \label{langeq}
\dot{x}(t)=-F\mathrm{sgn}(x)+ \sigma\xi_\alpha(t),
\end{equation}
where $x(t)=x$ denotes the equilibrium position at time $t$ and $^\bullet\equiv d/dt$. On the right-hand side of Eq. (\ref{langeq}), the first term is the negative potential gradient: $F$ is the magnitude of the downslide force, and $\mathrm{sgn}(x)$ is the signum function, which is $+1\,\,\forall\,+x$ and $-1\,\,\forall\,-x$. In the second term, $\sigma$ is the noise amplitude coefficient, i.e., the square-root of the noise variance $\sigma^2$, and $\xi_\alpha(t)$ is the noise function.

 When a particle is being kicked in the linear well, the noise variance $\sigma^2$ can be defined in terms of diffusion $D$ via the general solution to the diffusion equation $\partial_t\rho(x,t)=D\cdot\partial_x^2\rho(x,t)$:
\begin{equation} \label{gaussdist}
\rho(x,t)=\frac{1}{\sqrt{4\pi Dt}}\exp\left[-\frac{x^2}{2Dt} \right].
\end{equation}
From statistical mechanics, distributions are normalizable; Eq. (\ref{gaussdist}) is readily normalized, and thus, it defines a probability density function, where $\rho(x,t)=f(x,t)/Z_0$. Here, $f(x,t)$ is the function whereby the distribution has its profile, and $Z_0$ is the normalization factor:
\begin{equation}
Z_0=\int_{-\infty}^{\infty} f(x,t)dxdt.
\end{equation}
Since Eq. (\ref{gaussdist}) resembles a Gaussian pdf for Brownian noise, the variance is $\sigma^2={2Dt}$ with an average position value of $\langle x\rangle=0$, where
\begin{equation}
\langle a\rangle=\int_{-\infty}^{\infty}a(x,t)\cdot\rho(x,t)dxdt=\frac{1}{Z_0}\int_{-\infty}^{\infty}a(x,t)\cdot f(x,t)dxdt.
\end{equation}
Using the Gaussian pdf, the root-mean-square displacement for equilibrium noise is the noise amplitude: $x_{\mathrm{rms}}=\sigma=\sqrt{2Dt}$.

For non-equilibrium noise, there is no characteristic time $t$ for the linear down-slide back to return to equilibrium. Its counterpart $\tau$ is only dependent on the force in such a way that a large force leads to a short time to slide down, and vice versa. Therefore, $\tau\propto F^{-1}$; placing this time in place of $t$ in the noise amplitude provides the proportionality $\sigma\propto\sqrt{D/F}$. This indicates that the magnitudes of $F$ and $\sigma$, and how they are related, depend on the noise stability $\alpha$. 

\subsection{Numerics and Kick Generating}

The ability to computationally simulate an agitated particle in a potential well is possible. In this case, we have to consider that our particle is subject to random kicks along either side of the well while on its Brownian random walk. Such random possibilities of agitation require us to compute iterations within a set of ``timesteps," a timestep being a kick in either direction of the well at a singular $i$-th moment of time. For equilibrium noise ($\alpha=2$), a numerical simulation of Eq. (\ref{langeq}) is straight-forward to perform. We consider discretization so that the Langevin equation transforms from a differential equation to an equation dependent on iterations:
\begin{equation}
\Delta x_i=\left[F(x_i)+\sigma\xi_2(t_i)\right]\Delta t,
\end{equation}
where $\Delta x_i=x_{i+1}-x_i$, and $F(x_i)=-F\mathrm{sgn}(x_i)$ is the force term, which can either be $-F$ for random positive inputs or $+F$ for random negative inputs.

Over the course of numerical simulation and kick generation, the Gaussian noise contribution $\xi_2(t_i)$ at the $i$-th timestep is taken as $\xi_2(t_i)=\theta_{i,2}(\Delta t)^{-1/2}$, where $\theta_{i,2}$ is the $i$-th random number from a Gaussian distribution in a sequence of timesteps. Our discrete Langevin equation essentially represents an Euler scheme:
\begin{equation}\label{scheme}
x_{i+1}=x_i+\left[-F\mathrm{sgn}(x_i)+\sigma\frac{\theta_{i,2}}{\sqrt{\Delta t}}\right]\Delta t.
\end{equation}

Because the steepness of the V-shaped linear well is dependent on $F$, the Euler scheme could become inaccurate with an exceedingly large slope, i.e., a very large force magnitude. This can be rectified by letting $\Delta t$ be small, which in turn would make $\Delta x_i=x_{i+1}-x_i$ just as small. The smallness of $\Delta x_i$ would consider the discreteness of the iterations as continuous; thus, we look at the Langevin equation again.

For non-equilibrium noise ($1\leq\alpha<2$), Eq. (\ref{scheme}) does not gaurantee such success. Each of the large kicks of the L\'evy noise covers a significant amount of the slope of $V(x)$ in the course of one timestep. One might be tempted to make $\Delta t$ even smaller, but the redefinition of the Gaussian noise generator into a L\'evy noise generator is intuitive.

 With $\alpha=2$, $\xi_2(t_i)\Delta t=\theta_{i,2}\Delta t^{-1/2}$. With an arbitrary $\alpha<2$, consider that the $2$ in the Gaussian case becomes $\alpha$ in the non-equilibrium case. Therefore, for non-equilibrium noise, the random $i$-th number $\theta_{i,\alpha}$ shall be extracted from an $\alpha$-stable distribution, where our non-equilibrium Euler scheme becomes
\begin{equation}\label{noneqscheme}
\Delta x_i=-F\mathrm{sgn}(x_i)\Delta t+\sigma\theta_{i,\alpha}\Delta t^{1/\alpha}.
\end{equation}

To implement this non-equilibrium Euler scheme in \textit{Mathematica}, the commands ``RandomVariate'' and ``StableDistribution'' and a few lines of code readily generate a large set of $\alpha$-stable distributed numbers, thereby simulating Eq. (\ref{langeq}). This procedure is also valid for parabolic, circular, and double-parabolic potential wells  \cite{Yuvan2021, Yuvan2022ent, Yuvan2022sym, Bier2024}. 

\begin{figure}[h!]
\centering
\includegraphics[width=87mm]{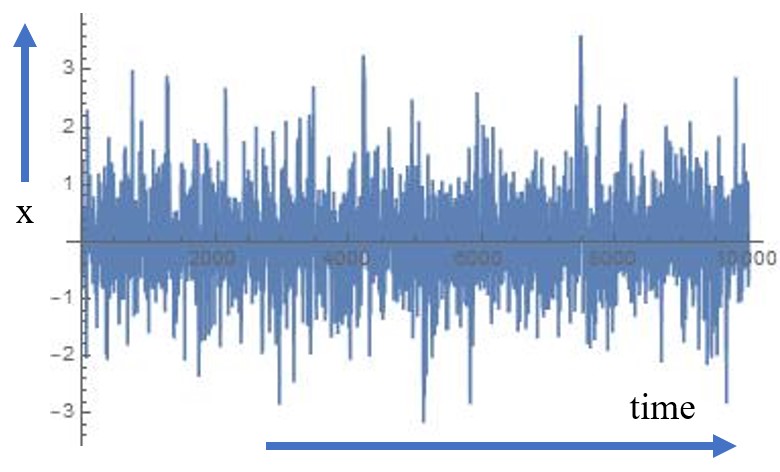}
\caption{\label{fig:vgauss} Gaussian noise ($\alpha=2$) in the V-shaped linear well via the Euler scheme. Due to microscopic reversibility, forward and reverse flows of time are indistinguishable.}
\end{figure}

\begin{figure} [h!]
\centering
\includegraphics[width=87mm]{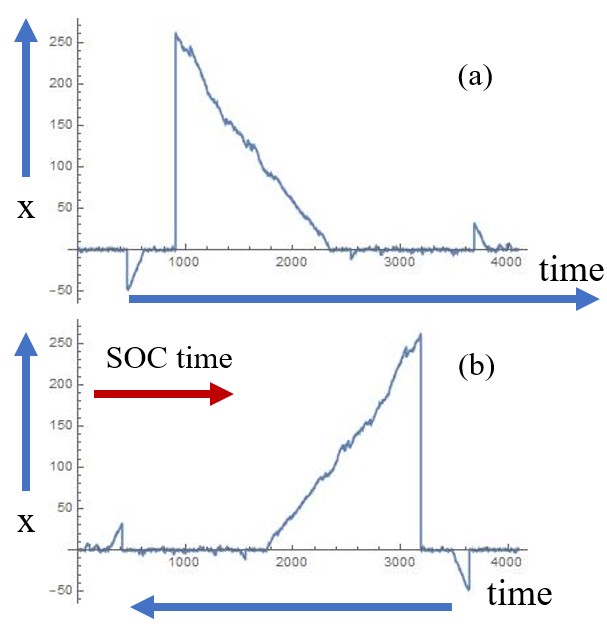}
\caption{ \label{fig:vlevy} L\'evy noise (using $\alpha=1.5$) in the V-shaped linear well via the Euler scheme. (a) A noise signal is simulated with a distinct flow of time (blue arrow). (b) By reversing the blue time arrow, the playback under a new flow of time (red arrow) reveals a characteristic SOC event.}
\end{figure}

As shown in Figures \ref{fig:vgauss} and \ref{fig:vlevy}, the L\'evy bath is characteristically Brownian under the conditions of $F>\sigma$ and $\alpha=2$, which presents time symmetry and upholds Lars Onsager's notion of ``microscopic reversibility" \cite{Onsager1931pr37, Onsager1931pr38}. Otherwise, if $F<\sigma$ and $\alpha<2$, the noise produced from perpetual flights and slow down-slides would result in a rough-edged sawtooth signal with broken time symmetry. Especially in Fig. \ref{fig:vlevy}, the noise signal produced by sudden kicks and slow down-slides can be horizontally reflected to display steady instability growth and sudden avalanches, i.e., a SOC process. 

\subsection{The Fractional Fokker--Planck Equation}

Since Eq. (\ref{langeq}) is an ordinary differential equation with stochastic inputs of $x$, it is possible to formulate an equivalent partial differential equation for how the pdf of the inputs $\rho(x,t)$, i.e., the distribution profile $f(x,t)$, evolves over time. 

For Gaussian noise, it is well known that the Langevin equation can be turned into the Fokker--Planck equation \cite{fp}. This equation traditionally has a first derivative with respect to $x$ to describe the drift terms and a second derivative with respect to $x$ that describes diffusion. For L\'evy noise, a noninteger $1<\alpha\leq2$ leads to the so-called fractional Fokker--Planck equation \cite{Jespersen1999}. In this case, instead of a second derivative with respect to $x$, we have a noninteger, i.e., a fractional derivative. 

The fractional Fokker--Planck equation corresponding to Eq. (\ref{langeq}) is 
\begin{equation} \label{ffpeq}
\frac{\partial}{\partial t}f(x,t)=F\frac{\partial }{\partial x}\Big[\mathrm{sgn}(x)f(x,t)\Big]+\sigma^\alpha\frac{\partial^\alpha }{\partial|x|^\alpha}f(x,t).
\end{equation}
For the steady state, $\partial_tf(x,t)=0$, which indicates that the distribution profile is only a function of $x$. From this condition, a stationary solution can be obtained from the following reduced equation:
\begin{equation} \label{lap}
\frac{\partial^{\,\alpha-1}}{\partial |x|^{\alpha-1}}f_{\mathrm{st}}(x)+\frac{F\mathrm{sgn}(x)}{\sigma^\alpha}f_{\mathrm{st}}(x)=0.
\end{equation}

Transformation spaces, such as Fourier $\mathcal{F}[.]$ and Laplace $\mathcal{L}[.]$, help solve impossible differential equations by turning derivatives into polynomials and convolutions into products. Transformations from one space to another involve integration, where
\begin{equation}
\begin{split}
&\mathcal{F}[f(x)]\equiv\tilde{f}(k)=\int_{-\infty}^{\infty}f(x)e^{-i2\pi kx}dx~~~\mathrm{and}\\
&\mathcal{L}[f(x)]\equiv\tilde{f}(s)=\int_0^\infty f(x)e^{-sx}dx,
\end{split}
\end{equation}
and the transformation variables ($k$ for Fourier and $s$ for Laplace) are in units of inverse length. After performing the transformation into one space and solving for the transformed function, an inverse transformation back into physical space must be taken: 
\begin{equation}
\begin{split}
&\mathcal{F}^{-1}[\tilde{f}(k)]\equiv f(x)=\int_{-\infty}^{\infty}\tilde{f}(k)e^{i2\pi kx}dk~~~\mathrm{and}\\
&\mathcal{L}^{-1}[\tilde{f}(s)]\equiv f(x)=\frac{1}{i2\pi}\int_{-i\infty}^{i\infty} \tilde{f}(s)e^{sx}ds.
\end{split}
\end{equation}

It is the Fourier transformation space that is widely used in noise analysis; filtering the noise realization over a long time series, i.e., over a large number of timesteps extracts the frequency values at which the noise sends out the largest signals. While the Fourier transformation space was utilized for the Hookian parabolic well $V(x)\propto x^2$ \cite{Yuvan2021}, where the scaled potential gradient convolution is $\partial_x[x\cdot f(x)]$ in normal space and $x\rightarrow\partial_k$ once Fourier transformed, it would not be as simple for the linear well profile. The Fourier transformation of the signum function is imaginary \cite{Burrows1990}:
\begin{equation}
\mathcal{F}\left[\mathrm{sgn}(x)\right]=\mathcal{C} \frac{2}{ik},
\end{equation} 
where $\mathcal{C}$ is the Cauchy principle value. Therefore, the Fourier transformation of Eq. (\ref{lap}) would become a complex number of the form $z=a+ib$ without solving the transformed distribution function $\tilde{f}(k)$. Alternatively, one can solve Eq. (\ref{lap}) in the Laplace transformation space.

Solving Eq. (\ref{lap}) in the Laplace space is well-studied for $1<\alpha<2$ \cite{Metzler2000, West2010, Lin2013, Medina2017, CapelasDeOliveria2019}. For $0<a<1$, where $a=\alpha-1$, the $a$-th order derivative of a function $f(x)$ with respect to $x$ can be expressed as
\begin{equation}
\frac{d^a}{dx^a}f(x)=\frac{1}{\Gamma(1-a)}\frac{d}{dx}\int_0^x\frac{f(x')}{(x-x')^a}dx',
\end{equation}
where $x>0$ and $\Gamma(.)$ is the Gamma function. Thus, the distribution function $f_{\mathrm{st}}(x)$ is readily solved \cite{Lin2013} to be
\begin{equation} \label{mlf}
f_{\mathrm{st}}(x)=\mathrm{E}_{\alpha-1}\left(-\frac{F}{\sigma^\alpha}|x|^{\alpha-1}\right),
\end{equation}
where 
\begin{equation}
\mathrm{E}_{a}(z)=\sum_{k=0}^{\infty}\frac{1}{\Gamma(ak+1)}z^k 
\end{equation}
is the Mittag--Leffler function. The boundary conditions of the $\alpha$ stability factor are applied to Eq (\ref{mlf}), to mathematically translate the Euler scheme under the two extremes of $\alpha$ via normalization.

\subsubsection{$\alpha=1$}

Using the summation definition of the Mittag--Leffler function, the summation with $a=0$  becomes a geometric progression sum for $|z|<1$:
\begin{equation}
\sum_{k=0}^\infty\frac{1}{\Gamma(1)}z^k=\frac{1}{1-z}.
\end{equation}
Since $z=-F/\sigma$ under $\alpha=1$, the obtained distribution is a flat line:
\begin{equation}
f_{\mathrm{st}}(x)=\frac{1}{(1+{F}/{\sigma})}.
\end{equation}
The geometric progression sets the condition that $|F/\sigma|<1$, which defines our well-understood L\'evy condition $F<\sigma$. This further explains the large flights and slow down-slides, which jeopardizes the distribution function from being normalizable:
\begin{equation}
Z_0=\int_{-\infty}^{\infty}\frac{dx}{(1+{F}/{\sigma})}=\infty,~~~\mathrm{or}~~~\frac{1}{Z_0}=0.
\end{equation}

A zero-flatline pdf tells us that there is no steady-state pdf for L\'evy noise. There is an equal probability for the particle to be anywhere in the linear potential well, which explains the likelihood of particles undergoing L\'evy flights while on their random walk. 

\subsubsection{$\alpha=2$}

The Mittag--Leffler function for $a=1$ is by definition the exponential function. Therefore, the specific distribution under Gaussian noise is a double-sided exponential decay:
\begin{equation}
\mathrm{E}_{1,1}\left(-\frac{F}{\sigma^2}|x|\right) = \exp\left(-\frac{F}{\sigma^2}|x| \right).
\end{equation}
which is easily normalizable:
\begin{equation} \label{res}
Z_0=\int_{-\infty}^{\infty}\exp\left(-\frac{F}{\sigma^2}|x| \right)dx=\frac{2\sigma^2}{F}.
\end{equation}

Unlike the previous case with L\'evy noise, where the particle can be anywhere within the potential well, the evaluation of $\langle x \rangle$ in the Gaussian case is an exact equilibrium point:
\begin{equation}
\langle x\rangle=\frac{F}{2\sigma^2}\int_{-\infty}^{\infty}x\cdot\exp\left(-\frac{F}{\sigma^2}|x| \right)=0.
\end{equation}
This preserves the quick relaxation of noise slightly out of equilibrium to the corner node at $x_0=0$.

\section{Normalizing the Mittag--Leffler Function for $1<\alpha<2$} \label{normal}

After analyzing the boundary conditions for $a=\alpha-1$, the stability region of $0<a<1$ must be addressed to ensure that the boundary conditions are met. The region is still considered to be non-equilibrium L\'evy noise, due to the Euler scheme demonstration in Figure \ref{fig:vlevy}. However, the question is whether L\'evy noise is normalizable above $a=0$. Because our Mittag--Leffler function takes the form of $\mathrm{E}_a(-t^a)$, where $t>0$ for $0<a<1$, the function was previously analyzed to be approximated with two asymptotic representations \cite{Mainardi2014}:
\begin{equation}
\mathrm{E}_a(-t^a)\sim
\begin{cases}
   e^0_a(t):= \exp\left[-\frac{1}{\Gamma(1+a)}t^a \right],~~ t\rightarrow 0\\\\
    e^\infty_a(t):=  \frac{1}{\Gamma(1-a)}t^{-a}\,,~~~~~~~~~~ t\rightarrow\infty 
     \end{cases} ,
\end{equation}
from which we obtain the inequality $e^0_a(t)\leq e^\infty_a(t)$ for sufficiently small and sufficiently large values of $t$.

The $t^a$ variable corresponds to the definition $F|x|^a/(\sigma^{a+1})$, and the L\'evy condition states that $|F/\sigma^{a+1}|<1$. As L\'evy noise is implied, it is intuitive to use the $t\rightarrow0$ definition for normalization over all $x$, provided that the smallness of $F/\sigma^\alpha$ truncates the Mittag--Leffler function to the ``small $t$'' asymptotic form. Calling $F/\sigma=c$ and $|x|/\sigma=|b|$, normalization is achieved by evaluating the integral over all $b$:
\begin{equation}
Z_0=2c^{1/a}\int_0^\infty \exp\left[\frac{-c|b|^a}{\Gamma(1+a)} \right]db=\frac{2\Gamma(1+1/a)}{\Gamma(1+a)^{1/a}}.
\end{equation}

With the obtained normalization factor, we define the pdf of noise signals within a V-shaped linear well under any stability factor $\alpha$:
\begin{equation} \label{alphapdf}
\rho_\alpha(x) =\frac{F}{2\sigma^\alpha}\frac{\Gamma(\alpha)^{1/(\alpha-1)}}{\Gamma\left(\alpha/(\alpha-1)\right)}\mathrm{E}_{\alpha-1}\left(-\frac{F}{\sigma^\alpha}|x|^{\alpha-1}\right),
\end{equation}
which preserves the $\alpha=1$ and $\alpha=2$ boundary conditions:
\begin{equation}
\begin{split}
&\rho_1(x)=\frac{F}{2\sigma}\frac{\Gamma(1)^{\infty}}{\Gamma\left(\infty\right)}\mathrm{E}_{0}\left(-\frac{F}{\sigma}\right)=0, \\
&\rho_2(x)=\frac{F}{2\sigma^2}\mathrm{E}_{1}\left(-\frac{F}{\sigma^2}|x|\right)=\frac{F}{2\sigma^2} \exp\left(-\frac{F}{\sigma^2}|x| \right).
\end{split}
\end{equation}

\begin{figure}[h!]
  \centering
    \includegraphics[width=87mm]{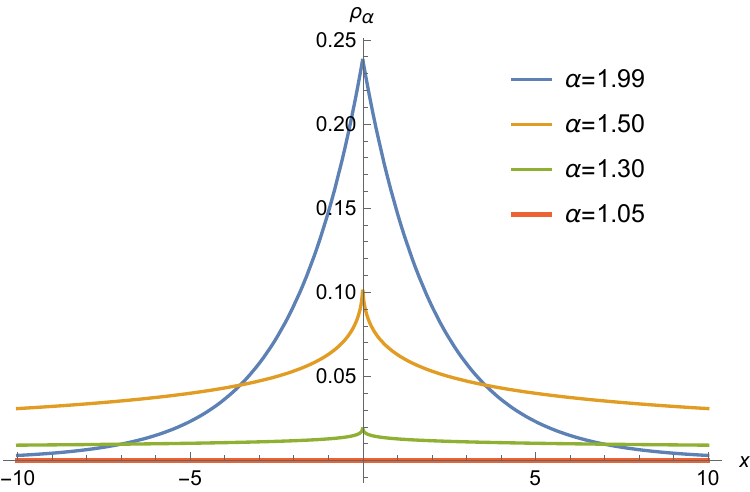}
    \caption{\label{fig:linwell} Non-equilibrium pdfs for specific values of $\alpha$ in the V-linear profile ($F=0.75$ and $\sigma=1.25$): $\alpha=1.99$ in blue, $\alpha=1.50$ in orange, $\alpha=1.30$ in green, and $\alpha=1.05$ in red.}
\end{figure}

Figure \ref{fig:linwell} shows the sample profiles of Eq. (\ref{alphapdf}) under various stability factors of $1<\alpha<2$ while using the parameter values $F=0.75$ and $\sigma=1.25$. As $F<\sigma$, we have a characteristic L\'evy noise distribution. Recall that $F>\sigma$ is only valid for $\alpha=2$ Gaussian noise. As seen in the figure, as $\alpha\rightarrow1$ starting from 1.99, the L\'evy noise pdf progressively becomes a dampened flatline distribution with a defined spike at $x_0=0$. That is, until it draws the smooth zero flatline for $\alpha=1.05$. The presence of the spike for the stability factors $1.30\leq\alpha\leq1.5$ depicted in the figure reveals that relaxation to equilibrium is achievable, even though increasing instability due to decreasing $\alpha$ makes it increasingly improbable. 

\section{Discussion}

The mathematics and computation procedures provided in Sections \ref{methods} and \ref{normal} are intended to be applied to complex systems known to exhibit self-organized criticality. This is dictated by the parameters $F,~\sigma$ and $\alpha$, where the relationship between the down-slide force $F$ and noise amplitude $\sigma$ (and the power of $\sigma$) depends on the noise stability $\alpha$. One may even be tempted to define an Ansatz that equates $F$ with $\sigma$ based on $\alpha$. However, this would lead to a fundamental issue in dimension analysis.

For non-physical ``L\'evian''  systems, i.e., complex and unstable systems, the SI units of physical force $F$ and noise amplitude $\sigma$ are irrelevant. As they are crucial to the Euler scheme defined as Eq. (\ref{scheme}), $F$ and $\sigma$ are input parameters for computationally reproducing datasets of the system in question. The comparison between the actual data and the computational data involves extracting the stability factor $\alpha$, thereby defining the corresponding pdf via Eq. (\ref{alphapdf}). For the parabolic well \cite{Yuvan2021}, $\alpha$ can be extracted by calculating the total number of backward and forward steps in the Euler scheme, i.e., $N_b$ and $N_f$, dividing the difference of the steps by its sum to define the parameter $r$, and subjugating it into an algorithm provided in Appendix B in Ref. \cite{Yuvan2021}. This algorithm can be applied to the linear well Euler scheme.

\section{Conclusion}

In this short study, I addressed two key aspects of non-equilibrium noise analysis. The first is an Euler iteration scheme defined as Eq. (\ref{scheme}), which is derived by discretizing the Langevin equation (Eq. [\ref{langeq}]). These equations depend on the scaling inequality between the noise amplitude $\sigma$ and the constant downslide force $F$; how the inequality is defined is inherently based on the stability of the noise $\alpha$. L\'evy noise of stability $1<\alpha<2$ is characterized by sudden kicks due to external agitation with a slow down-slide to achieve equilibrium. Gaussian noise of stability $\alpha=2$ is characterized by the reverse case, i.e., small jitter events that quickly slide back down to equilibirum. 

The likelihood of achieving equilibrium under any $\alpha\in(1,2]$ is described by the probability density function, which is the second key aspect offered in this short study. The derivation of the pdf, defined as Eq. (\ref{alphapdf}), is possible by obtaining the fractional Fokker--Planck equation (Eq. [\ref{ffpeq}]) and its steady state reduction (Eq. [\ref{lap}]) from Eq. (\ref{langeq}). Using the Laplace transformation space and asymptotic forms of the Mittag--Leffler function (Eq. [\ref{mlf}]), one can see how the stability factor changes the profile and behavior of the L\'evy noise pdf. One can use the pdf profile under any $\alpha\rightarrow1$ to determine how increasingly likely it is to have L\'evy flights and how increasingly improbable it is to achieve equilibrium.

Eqs. (\ref{scheme}) and (\ref{alphapdf}) are applicable to systems that exhibit self-organized criticality. While such systems exist in the unnatural world, SOC applications to physics and the rest of the natural world, beyond what was mentioned in the introduction, are open for exploration. 

\subsection*{Dedication}
This short study is dedicated to Per Bak (1948-2002), one of the founders of and the strongest advocate for the concept of self-organized criticality.

\end{document}